# On PROGRESS Operation

*How to Make Object-Oriented Programming System More Object-Oriented (DRAFT)*

Evgeniy Grigoriev
*www.RxO project.com*
Grigoriev.E@gmail.com

Abstract: A system, which implements persistent objects, has to provide different opportunities to change the objects in arbitrary ways during their existence. A traditional realization of OO paradigm in modern programming systems has fundamental drawbacks which complicate an implementation of persistent modifiable objects considerably. There is alternative realization that does not have these drawbacks. In the article the PROGRESS operation is offered, which modify existing object within an existing inheritance hierarchy.

## 1 INTRODUCTION

OO programming paradigm (Booch,1991) claims to be the best and most natural way to model the real world in information systems. Existing OO programming tools are the result of years of von Neumann machines programming systems evolution. Organization of target ALM-machines had and has a significant impact on abilities and features of existing OO languages. An important feature of the machines is the using of addressable linear memory (further, ALM). As shown in (Grigoriev, 2012), core features of the addressable linear memory make an implementation of persistent modifiable objects a very complex task.

An obvious drawback is that the AML itself is not persistent usually. But it's not the only negative feature. Let us consider other ones. First feature is that both links between objects and internal structures of the objects are mapped into a single address space. Second one is that unidirectional address pointers are the only possible ways to link memory areas. These features together are critical when modifiable structures are tried to be implemented. A memory area allocated to an object data is limited by the areas allocated to other objects. An attempt to modify an object structure, which increases the corresponding memory area, requires reallocation of the area in the memory. All address pointers referencing to the reallocated memory area have to be changed to keep links existing in the system. But there is no a way to track existing address pointers because of their unidirectionality. If a memory area is reallocated, the links will be lost.

Of course, all these difficulties can be got round by sophisticated programming. A huge amount of very different technologies, approaches, patterns, environments etc. exist, which try to implement persistent modifiable objects. A result of such programming generally looks like an attempt to avoid or to hide some negative features of used ALM-machines.

Alternative way to avoid totally these features is to avoid ALM-machines themselves.

## 2 OTHER MACHINES

ALM-machines are not the only target machines for OO programming systems. In (Grigoriev, 2012) an object-oriented translator is described which uses a relational database as a target machine (R-machine). This possibility is based on the fact, that today relational DBMS fully correspond to the concept of the target machine (Pratt and Zelkowitz, 2001). They are programmable data systems which can create, save, and execute a command sequences on relational variables and on their values. Of course, they are virtual machines, but from formal standpoint this fact has no matter. In comparison

with the ALM-machines, the R-machines have the following features:
- associative principles of memory organization;
- persistent memory;
- ability to manipulate with groups of values by means of set operations;
- formal foundation (relational data model).

The proposed approach "OO translator for R-machine" was used to create "RxO system" prototype (Grigoriev, 2013-1), which fully combines the core properties of object-oriented languages and relational DBMS. It is based on a formal possibility to convert the description of complex object-oriented data structures and operations on these data, into a description of relational structures and operations on last ones (Grigoriev, 2013-2),.

Data in this system are described as a set of persistent objects and represented as an orthogonal set of relations. The system is managed by commands of non-procedural language. The commands are used to create and to change classes, to create persistent objects and change their state, to get data about the objects state (inc. by means of ad-hoc queries), to execute object methods, to manipulate with groups of the objects.

The above features of the target R-machines affect properties of the source OO language appreciably, giving the possibilities that are not present in the traditional object-oriented languages. Associative memory of R-target machine has no principal drawbacks that prevent easy implementation of persistent modifiable objects. Partially, this ability is demonstrated in the "RxO system" prototype; new classes can be added on-the-run into a set of existing ones in working models (inc. by mean of multiply inheritance), implementations of attributes and methods can be changed in existing objects

Fundamentally, object interfaces can be changed too in different ways. This feature can be realized in two different operations, which is going to be implemented into next version of RxO system.

The first operation is used to change a class specification class, for example, to add new attribute or method. This operation affects all objects of the class by changing their interface. Its meaning is clear, so we will not dwell on it.

The second operation leaves all existing specifications unchanged but puts existing object into other group of objects defined by a class or by a role. It means that interface and implementation of the objects is changed as it is defined for group objects, to which the object will belong after operation.

Of course, the ability to put any existing object in any class has a little sense. In addition, such ability will evidently contradict to a static typing that is implemented in RxO system. However, limited versions of such operation exist, which do not contradict to the static typing. Interestingly, it is precisely the limitation, which makes this operation meaningful as a way to create more perfect information model of object domains.

## 3 OBJECTS IN PROGRESS

Consider the information system, which is an information model of some firm. The firm staff consists of employees and some of the employees are managers. Data in the system are describes as a set of persistent objects.

### 3.1 Progress in Inheritance Hierarchy

Let's take the classic example used often to demonstrate the inheritance principle, where `Manager` class is a subclass of `Employee` class.

```
CREATE CLASS Employee
{
    …
}

CREATE CLASS Manager
EXTENDS Employee
{
    …
}
```

Accordingly, objects of `Employee` class exist in the system among others objects, and some of them belong to `Manager` class. All these objects can be referenced from other objects of the system, and the inherited class existence has no meaning in some cases of the referencing; for example, `Library` object referencing all the `Employee` objects equally to track issued books, even if some of the `Employee` objects are `Manager` objects.

The world is inconstant and a time has come when some employee is promoted to a manager position. It's important that the employee has not become a different person. For example, its relationship with the firm library has not changed in this moment, but the promotion is very important for HR department. Remaining the same, the employee has acquired a new quality, entered into a new group. It has progressed.

The simplest way to reflect this situation in the information system is an operation that shifts an object of `Employee` class up within the inheritance hierarchy, which has already been defined. Let us call this operation as `PROGRESS`. This operation keeps both OID of applied object and its interface, described in the parent class, unchanged. Therefore all existing references to the objects stay valid. At the same time, the object gets new interface elements as they are defined in the child class. So, after the object was progressed, it can be used by other objects of the system as an object of the derived class. In RxO system, which implements a principle "class is a stored set of objects", the object also will be available as an element of the child class.

The `PROGRESS` variant of putting of object in other class doesn't reduce the system reliability, which is achieved by static typing, because the old object type remains unchanged by this operation.

The implementation of the interface, defined for objects of the parent class, can be redefined in the child class. The `PROGRESS` operation can be used with a special method (re-constructor) to transform the object to make it corresponding to the new implementation. This re-constructor can take parameters to add new data into the object during the transformation. So, the operation syntax can look like

```
PROGRESS someEmployee
    TO Manager(parameter, ….)
```

A combination of the `PROGRESS` operation and an ability to inherit on-the-run classes of an information models provides a simple and logical way to develop the model, in order to reflect changes of an object domain. If it's necessary, a class can be inherited by a new-created class, which has new specification elements and/or changes implementations of existing elements. Then, objects of the class can be progressed to the new child class.

## 3.2 Progress in Roles

Other interpretation of "Employee to Manager" example can be offered, where "a manager" is considered just as one of roles, which are possible for an employee.

A role is applied to a class. Like a class, a role can have both attributes and methods which have to be implemented before role can be used. Also it can have special method (role constructor) to build a role data from the applied object or/and from taken parameters. In RxO system the role defines a set of objects which the role has been applied to. Speaking generally, a role definition is very similar to inherited class definition. The only difference is that a role cannot re-implement an applied class.

Roles in RxO system look similar to class interfaces available in some traditional OO languages like Java and C#. As opposed to the interfaces, roles can have attributes

An advantage of the roles is clear when an object can have different independent roles. For example some employee can be a manager for other employees or/and a mentor for new-coming employee. These two roles are independent.

```
CREATE CLASS Employee
{
    …
}

CREATE ROLE Manager FOR Employee
{
    …//the same as in inherited class
}

CREATE ROLE Mentor FOR Employee
{
    …
}
```

After the two roles was created and implemented, some `Employee` object can be progressed to both these roles.

```
PROGRESS someEmployee
    TO Manager(parameter, ….)

PROGRESS someEmployee
    TO Mentor(parameter, ….)
```

Now the object can be used in the both new roles by other objects of the system. In RxO system, which implements a principle "class is a stored set of objects", the object also will be available as an element of both object groups defined by the roles.

This situation can be hardly described by usual class inheritance because two different child classes would define two non-overlapping subsets of objects. Subsets, defined by independent roles, can overlap.

A role can be applied to other role. For example `Top` role can be defined for `Manager` one.

```
CREATE ROLE Top FOR Manager
{
    …
}
```

Now the `Employee` object having `Manager` role can get new `Top` role.

```
PROGRESS someManager
    TO Top(parameter, ….)
```

At that all other independent roles will stay unchanged.

A combination of the `PROGRESS` operation and an ability to create on-the-run new roles provides other way to develop the model. If it's necessary, a new role can be created for existing class or role. Then, existing objects can be progressed to the new role.

## 4   CONCLUSIONS

Persistent objects are meaningful only if they are truly modifiable, because of modeled object domain inconsistence. Here the word "modifiable" has the widest sense which includes at least the following items:

- an ability to change the state of the object;
- an ability to change on-the-run the class implementation, that changes behavior of all its objects;
- an ability to change on-the-run the class specification, that changes interface of all its objects;
- an ability to put an existing object in child class or in applied role, that can change the state and the behavior of the object and add new elements to its interface.

In systems which claims to be the best way to model the real world, all these abilities have to be equally and easily accessible. But most of traditional OO languages have only the first ability implemented as a basic one. Other three abilities require a non-trivial programming or are unavailable at all.

As an example, obvious `PROGRESS` operation can be hardly implemented in existing OO programming systems for ALM-machine. Perhaps, this is why such operations are practically unknown (similar operations exist in very rare classless languages, for example, NewtonScript). This situation clearly demonstrates an impact of architecture of habitual target ALM-machines on OO languages and also on understanding of OO paradigm itself.

A reason of the described drawback is not the OO paradigm itself but only its usual implementation. Other implementations of the paradigm can exists which do not have such drawbacks. An example of such implementation is the "RxO system" prototype.